\documentclass[prl,preprint,showpacs]{revtex4}
\usepackage{amsmath}
\usepackage{graphicx}
\usepackage{dcolumn}
\usepackage{bm}
\usepackage{float}

\newcommand{\beq}{\begin{equation}}
\newcommand{\eeq}{\end{equation}}
\newcommand{\bea}{\begin{eqnarray}}
\newcommand{\eea}{\end{eqnarray}}
\newcommand{\beano}{\begin{eqnarray*}}
\newcommand{\eeano}{\end{eqnarray*}}

\begin{document}

\title{Hyperfine interaction induced critical exponents in the quantum Hall
effect }
\author{V. Kagalovsky$^{1}$ and \frame {I. Vagner}$^{2,3}$}
\date{\today}

\begin{abstract}
We study localization-delocalization transition in quantum Hall systems with
a random field of nuclear spins acting on two-dimensional (2d)
electron spins via hyperfine contact (Fermi) interaction. We use
Chalker-Coddington network model, which corresponds to the projection onto
the lowest Landau level. The inhomogeneous nuclear polarization acts on the
electrons as an additional confining potential, and, therefore, introduces
additional parameter $p$ (the probability to find a polarized nucleus in the vicinity 
of a saddle point of random potential) responsible for the change from quantum to
classical behavior. In this manner we obtain two critical exponents
corresponding to quantum and classical percolation. We also study how
the 2d extended state develops into the one-dimensional (1d) critical state.
\end{abstract}


\affiliation{$^{1}$Sami Shamoon College of Engineering, Beer-Sheva,
Israel.\\ {
\small {\textit{$^{2}RCQCE-$Research Center for Quantum
Communication Engineering,}} }\\
{\small \textit{Holon Institute of Technology, 52 Golomb St., Holon 58102,
Israel} }\\
$^{3}${\small Cavendish Laboratory, \ \ Cambridge University, \ Madingley
Road, Cambridge CB3, OHE, UK.}}

\maketitle

Celebrated quantum Hall effect (QHE) is realized in a 2d 
electron gas subjected to a strong perpendicular magnetic field and a
random potential \cite{KDP80,Prange}. The uniqueness of this
phenomenon is in high precision of the plateaux in the Hall component and
very rich physics in the interplateau transitions \cite{82Iord}. Here we
will study the influence of the nuclear spin fields \cite{VM88} on the
critical exponents in QHE.

The rich physics of the random potential in quantum Hall systems could be
roughly divided into spin independent and spin dependent (spintronics)
electron scattering processes. Magnetic impurities perturb the QHE transport
very strongly and will not be considered here. Recently sharply growing
attention was attracted to the physics of the hyperfine interactions in
the~QHE. It was suggested theoretically \cite{VM88} and observed
experimentally \cite{Berg90,04Yusa} that the underlying nuclear spin
structure can provide the microscopic information on the 2d electron wave
functions and provide strong influence on the precision and other parameters
of a QHE system.

The interaction between electron and nuclear spins in heterojunctions
under QHE conditions is due, usually, to the \textit{hyperfine Fermi contact}
interaction \cite{AbragBk61,VM88}. This interaction is represented by the
Hamiltonian: 
\begin{equation}
\hat{H}_{int}=-\gamma _{n}\hbar \vec{I}_{i}\cdot \vec{H}_{e},  \label{Hint1}
\end{equation}%
where $\gamma _{n}$ is the nuclear gyromagnetic ratio, $\vec{I}_{i}$ is the
nuclear spin and $\vec{H}_{e}$ is the magnetic field on the nuclear site, produced
by electron orbital and spin magnetic moments: 
\begin{equation}
\vec{H}_{e}=-g\beta \sum_{e}{\frac{8\pi }{3}}\hat{s}_{e}\delta \left( \vec{r}%
_{e}-\vec{R}_{i}\right) .  \label{He1}
\end{equation}%
Here $\vec{r}_{e}$ is the electron radius-vector, $\hat{s}_{e}$ is the
electron spin operator, $\beta =e\hbar /m_{0}c$ is the Bohr magneton, $g$
is the electronic $g$-factor and $\vec{R}_{i}$ is the nucleus radius-vector. 

It follows from Eqs.\ (\ref{Hint1}) and (\ref{He1}), that once the nuclear spins are
polarized, i.e. if $\left\langle \sum_{i}\vec{I}_{i}\right\rangle \neq 0$
, the charge carriers spins feel the effective, time-dependent hyperfine field $%
B_{hf}=B_{hf}^{o}\exp \left( -t/T_{1}\right) $ ($T_1$ is a nucleus relxation time)
which lifts the spin
degeneracy even in the absence of external magnetic field. In GaAs/AlGaAs
one may achieve the spin splitting due to hyperfine field of the order of
the Fermi energy \cite{Berg90,04Yusa}.The inhomogeneous nuclear
polarization acts on the electrons as additional (to the scalar potential of
the impurities) confining potential $V_{hf}=-\mu _{B}B_{hf}$ \cite{NSPI}.

The nuclear spin polarization, once created, remains finite for
macroscopically long times. Intensive experimental studies \cite{Berg90,04Yusa} 
of this phenomenon in QHE systems have provided a more
detailed knowledge on the hyperfine interaction between the nuclear and
electron spins in heterojunctions and quantum wells. It was observed that
the nuclear spin relaxation time is rather long (up to 10$^{3}\sec $) and
the hyperfine field acting on the charge carriers spins is extremely high,
up to 10$^{4}G$ \cite{Berg90}. The nuclear relaxation time depends strongly
on the vicinity to the impurity and its sign \cite{82Iord}. The presence of
the impurity (long range potential) provides the necessary energy
conservation in the spin-exciton creation process leading to the nuclear spin
relaxation. We can, therefore, expect the following scenario: nuclear spins
being polarized by some external field will then relax differently depending
on whether they are close to maxima or minima of the scalar potential
created by impurities. Therefore, they should affect strongly tunneling of
electrons through saddle-point potential.

When random potential varies smoothly (its correlation length is much larger
than the magnetic length as, e.g., in GaAs heterostructures) a semiclasscial
description becomes relevant: electrons move along the lines of constant
potential. When two equipotential lines come close to each other (near a
saddle point) tunneling is feasible. In this paper we investigate how this
picture will be affected by strong nuclear polarization. We find that
scaling of the localiztion length is modified (Eq. (8)), which is the main
result of this work.

In the network model \cite{rchalk}, electrons move along unidirectional
links forming closed loops in analogy with semiclassical motion on contours
of constant potential. Scattering between links is allowed at nodes in order
to map tunneling through saddle point potentials. Propagation along links
yields a random phase $\phi $, thus links are presented by diagonal matrices
with elements in the form $\exp (i\phi )$. Transfer matrix for one node
relates a pair of incoming and outgoing amplitudes on the left to a
corresponding pair on the right; it has the form 
\begin{equation}
\mathbf{T}=\left( 
\begin{array}{cc}
\sqrt{1+\exp (-\pi\epsilon )} & \exp (-\pi\epsilon /2) \\ 
\exp (-\pi\epsilon /2) & \sqrt{1+\exp (-\pi\epsilon )}%
\end{array}%
\right) .  \label{first}
\end{equation}

In order for a system to be invariant, on average, under $90^{\circ }$
rotation the transmission and reflection at the next neighbor node are
interchanged, i.e. the transfer matrix has the same as in Eq.\ (\ref{first})
form with a parameter $\epsilon ^{\prime }=-\epsilon $ \cite{rchalk}. In
order to obtain this relation one simply interchanges $Z_{3}$ and $Z_{4}$
(see Fig. 1) and brings a new transfer matrix to the form of Eq.(\ \ref%
{first}) . We therefore describe scattering at the nodes indicated in Fig. 1
by circles with transfer matrix $\mathbf{T}(\epsilon )$ and at the nodes
indicated by boxes with $\mathbf{T}(-\epsilon)$.

The node parameter $\epsilon $ is a relative distance between the electron
energy and the barrier height. It is related to the physical quantities
descibing the system

\begin{equation}
\epsilon\equiv (E-(n+\frac{1}{2})E_2-V_0)/E_1,  \label{eqeps}
\end{equation}
where $E_1$ measures the ratio between saddle-point paramters and magnetic
field, $E_2$ is a distance between Landau levels at strong magnetic fields,
and $V_0$ is a reference point of a scalar potential (see [%
\onlinecite{rfertig}] for details). It is easy to see that the most
"quantum" case (equal probabilities to scatter to the left and to the right)
is at $\epsilon =0$, in fact numerical calculations \cite{rchalk} show that there is an
extended state at that energy. Numerical simulations on the network model
are performed in the following way: one studies system with fixed width $M$
and periodic boundary conditions in the transverse direction. Multiplying
transfer matrices for $N$ slices and then diagonalizing the resulting total
transfer matrix , it is possible to extract the smallest Lyapunov exponent $%
\lambda$ (the eigenvalues of the transfer matrix are $\exp(\lambda N)$). The
localization length $\xi_M$ is proportional to $1/\lambda$. Repeating
calculations for different system widths and different energies it is
possible to show that the localization length $\xi_M$ satisfies a scaling
relation 
\begin{equation}
\frac{\xi_M}{M} =f\left(\frac{M}{\xi (\epsilon )}\right) .  \label{second}
\end{equation}
In the QHE the thermodynamic localization length $\xi (\epsilon )\sim
|\epsilon |^{-\nu}$ and $\nu=2.5\pm 0.5$. This is the main result \cite%
{rchalk} and it is in a good agreement with experimental data for spin-split
resolved levels \cite{rkoch}, numerical simulations using other models \cite%
{rhuck1} and semiclassical argument \cite{rmyl,weeuro} that predicts $\nu
=7/3$.

It is possible to model classical percolation using CC model as well. It was
shown \cite{kivel} that when the relative height of the barriers fluctuate
in the infinite range, the percolation becomes classical (no tunneling is
allowed) and classical percloation exponent $\nu _{cl}=4/3$ is retrieved. On
the other hand, when the fluctuations are finite, their width acts as
irrelevant parameter \cite{rhuck,reast,moith} and does not affect $\nu $. 

In the present work we modify CC model in the following way. We expect that
the presence of a polarized nucleus near a saddle point of the scalar
potential will modify a tunneling parameter $\epsilon$ in Eq. (3) by
changing $V_0$ to $V_0\pm V_{hf}$. More, we also expect that due to
different relaxation rates (in the vicinity of impurities of different signs)
the following scenario can be realized: nuclei situated near different types
of saddle point (nodes of the model) will be polarized in opposite
directions, breaking, therefore, isotropy of the system. We model this
situation by introducing a parameter $0\leq p<1$ describing the probability
that there is a polarized nucleus near particular saddle point. Due to the
effect of high hyperfine fields descibed above, we, as a rough
approximation, can expect that the barrier becomes "infinite", i.e. the
transfer matrix at the node is now a unit matrix. On the model language it
means that the quasiparticle stays on the same horizontal link (see Fig. 1),
and isotropy of the model is therefore broken. Obviously, when $p=1$ a 2d
system is broken into $M$ one-dimensional chains, and, due to the fact that
there is no backscattering, all states are extended independent on energy $%
\epsilon$ and system width $M$. We, therefore, expect the smallest Lyapunov
exponent $\lambda =0$, in contradistinction to the "ordinary" 2d extended
state, where $\lambda$ is finite, and infinite thermodynamic localization
length is recovered only after finite size scaling. In this sense $p=1$ case
is close to a 1d metal found for a dirty superconductors with broken
time-reversal and spin-rotational symmetries \cite{weprb}.

\begin{figure}[tbp]
\par
\begin{center}
\includegraphics[bb=97 525 368 794,scale=0.7]{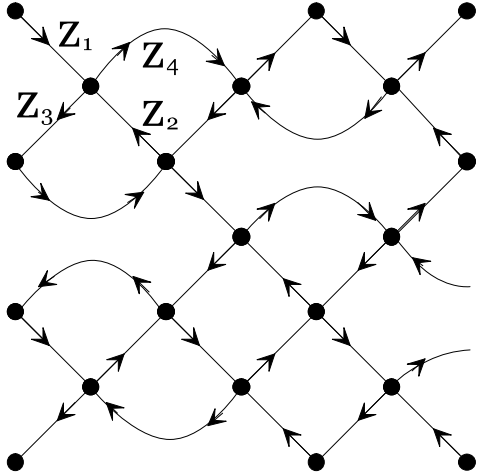}
\end{center}
\caption{Network model with missing nodes}
\end{figure}

Before we present numerical results, let us discuss the possible form for
the scaling of the renormalized localization length. Now, when we have "wiped
out" on average a fraction $p$ of the nodes, a quasiparticle should travel
larger distance (times $1/(1-p)$) in order to experience the same number of
scattering events. Therefore, naively, one would expect that the effective
system width is now $M(1-p)^{-1}$ and the scaling is

\begin{equation}
\frac{\xi_M}{M} =(1-p)^{-1}f\left(\frac{M}{\xi (\epsilon )}\right) .
\label{second}
\end{equation}

On the other hand, we should take into account that the "missing" node
actually does not allow the quasiparticle to propagate in the transverse
direction (we have chosen the system in such way that, if there is no
scattering, the quasiparticle stays on the same \textit{horizntal} link).
Usually for CC model and its generalizations the typical value of the
renormalized localization length for the extended state is of the order of $1$,
meaning that in the extended state the quasiparticle is able to traverse the
system of the width $M$. Therefore, in the present situation we could expect
even larger value of $\xi_M$ in the extended state, i.e. $(1-p)^{-\nu} $
dependence with $\nu >1$.

In order to find both critical exponents we start by studying a $p$%
-dependence for $\epsilon =0$, corresponding to the development of a 2d
extended state into a 1d extended state. The results for system widths $%
M=16,32,64$ are presented on Fig. 2, allowing the following fit

\begin{equation}
\frac{\xi_M}{M} =(1-p)^{-1.3}f(0) ,  \label{third}
\end{equation}
where $f(0)$ is the value of the renormalized localization length in the
extended state $\epsilon =0$ for the standard CC model ($p=0$). This value
for the critical exponent is suspiciously close to the classical percolation
exponent $\nu_{cl}=4/3$. We also show visibly worse fit of the data with the
"naive" critical exponent $\nu =1$.

\begin{figure}[tbp]
\par
\begin{center}
\includegraphics[bb=9 11 290 227,scale=0.7]{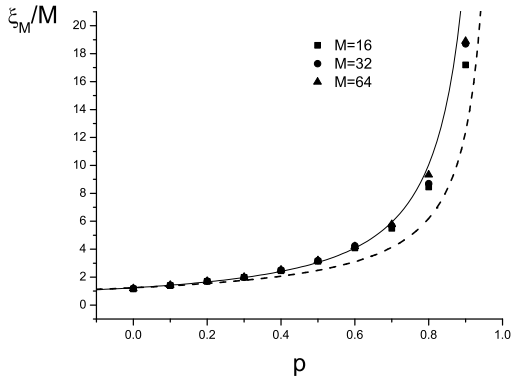}
\end{center}
\caption{Renormalized localizaton length at critical energy $\protect%
\epsilon =0$ as function of the fraction of missing nodes $p$ for different
system widths. Solid line is the best fit $1.24(1-p)^{-1.3}$. Dashed line is
the fit with "naive" exponent $\protect\nu =1$.}
\end{figure}

We next use the value $\nu$ found in Fig. 2 and study numerically
renormalized localization length for various $\epsilon \neq 0$ and $p<1$.
All our data collapse on one curve with abscissa in the form $M/\xi
(\epsilon )$ where thermodynamic localization length diverges as $%
\xi\sim\epsilon^{-\nu_{q}}$ with quantum percolation exponent $\nu
_{q}\approx 2.5$. The results of the scaling are presented on Fig. 3.

\begin{figure}[tbp]
\par
\begin{center}
\includegraphics[bb=17 17 293 224,scale=0.7]{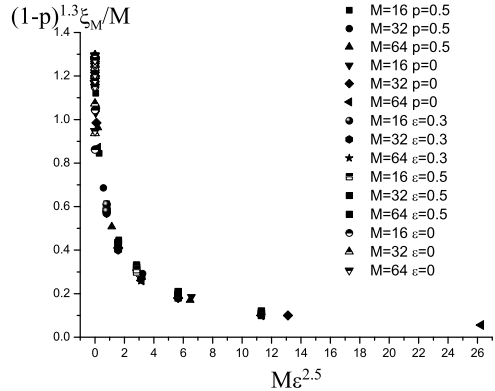}
\end{center}
\caption{Data collapse for all energies $\protect\epsilon $, system widths $%
M $ and all fractions $p\neq 1$ of missing nodes.}
\end{figure}

We argue that one can understand the appearance of the classical percolation
exponent in Eq. (7) by considering a quasiparticle on the standard CC model
deeply into the localized regime. In this case localization length $\xi_M$
is $M$-independent, meaning that a quasiparticle does not "feel" the
boundaries of the system, and its thermodynamic localization length $\xi
=\xi_M$. Therefore, a quasiparticle travels on the perimeter of the
classical cluster of the typical size $\xi$. Then by increasing the fraction 
$p$ of the missing nodes, we increase the size of the classical cluster,
actually making it infinite as $p$ approaches $1$. Therefore, $(1-p) $ acts
as energy in the classical percolation problem, explaining the value $%
1.3\approx 4/3$.

Finally, all numerical data we have obtained supports the following scaling
relation 
\begin{equation}
\frac{\xi _{M}}{M}=(1-p)^{-\nu _{cl}}f(M\epsilon ^{\nu _{q}}),
\end{equation}%
We stress that this is the first result for the network models to
produce both quantum and classical percolation exponents form the same data.
To summarize, we have studied the influence of nuclear spins on the
localization-delocalization transition in quantum Hall systems. We have
found that the fraction $p$ of polarized nuclei acts as a relevant
parameter, leading to a new scaling relation for the localization length
(Eq. (8)).

One of us (V. K.) appreciates valuable discussions with Alexander
Mirlin,Yuval Gefen, Baruch Horovitz and Yshai Avishai. I.V. is gratefull to
Mike Pepper, Cavendish, for hospitality during the summer 2005, when part of
this paper was accomplished. I.V. acknowledge the EuroMagNET of FP6,
RII3-CT-2004-506-239.

{\it Note added}. When this paper was already submitted my coauthor Professor Israel Vagner has died
after a lengthy and courageous battle with cancer.

\end{document}